\documentclass[conference]{IEEEtran}
\renewcommand\IEEEkeywordsname{Keywords}

\usepackage[pdftex]{graphicx}
\usepackage{float}

\usepackage[autostyle]{csquotes}
\usepackage{fancyhdr}

\ifCLASSINFOpdf
\else
\fi
\usepackage{amsmath}
\ifCLASSOPTIONcompsoc
  \usepackage[caption=false,font=normalsize,labelfont=sf,textfont=sf]{subfig}
\else
  \usepackage[caption=false,font=footnotesize]{subfig}
\fi

\begin{document}

		\pagestyle{fancyplain}
		\fancypagestyle{plain}{}
		\fancyhf{} 
		\fancyhead[C]{\textit{Accepted as conference paper at the ISCAS 2020}}
		\fancyfoot[C]{\textit{{\small \textcopyright 2020 IEEE.  Personal use of this material is permitted.  Permission from IEEE must be obtained for all other uses, in any current or future media, including reprinting/republishing this material for advertising or promotional purposes, creating new collective works, for resale or redistribution to servers or lists, or reuse of any copyrighted component of this work in other works.}}}
%		\fancyfoot[L]{\leftmark}
%		\fancyfoot[R]{\thepage}
		\renewcommand{\footrulewidth}{0.1mm}
		\renewcommand{\headrulewidth}{0.1mm}

\title{Heartbeat-Based Synchronization Scheme for the Human Intranet: Modeling and Analysis}

\author{\IEEEauthorblockN{Robin Benarrouch\IEEEauthorrefmark{1}\IEEEauthorrefmark{2}\IEEEauthorrefmark{3},
Ali Moin\IEEEauthorrefmark{3},
Flavien Solt\IEEEauthorrefmark{4}\IEEEauthorrefmark{1}\IEEEauthorrefmark{3}, Antoine Frappé\IEEEauthorrefmark{2},
Andreia Cathelin\IEEEauthorrefmark{1},
Andreas Kaiser\IEEEauthorrefmark{2}\\ and
Jan Rabaey\IEEEauthorrefmark{3}}

\IEEEauthorblockA{\IEEEauthorrefmark{1}STMicroelectronics, Crolles, France, Email: robin.benarrouch@st.com}
\IEEEauthorblockA{\IEEEauthorrefmark{2}Univ. Lille, CNRS, Centrale Lille, Yncréa ISEN, Univ. Polytechnique Hauts-de-France, UMR 8520 - IEMN, Lille, France}
\IEEEauthorblockA{\IEEEauthorrefmark{3}BWRC, University of California Berkeley, Berkeley, CA, USA}
\IEEEauthorblockA{\IEEEauthorrefmark{4}Ecole Polytechnique, IP Paris, Palaiseau, France}}


\maketitle
\thispagestyle{fancyplain}

\pagestyle{empty}
\begin{abstract}
	Sharing a common clock signal among the nodes is crucial for communication in synchronized networks.
%	One way to enable communication within a network is to share a clock signal for synchronization. 
	This work presents a heartbeat-based synchronization scheme for body-worn nodes. The principles of this coordination technique combined with a puncture-based communication method are introduced. Theoretical models of the hardware blocks are presented, outlining the impact of their specifications on the system. Moreover, we evaluate the synchronization efficiency in simulation and compare with a duty-cycled receiver topology. Improvement in power consumption of at least 26\% and tight latency control are highlighted at no cost on the channel availability. \newline
\end{abstract}



\IEEEpeerreviewmaketitle

\section{Introduction}\label{sec_Intro}
	The Human Intranet, introduced in \cite{rabaey2015human}, is a human body-dedicated network.
%	It focusses on describing and identifying the challenges of deploying such a system
	As part of Wireless Body Area Network (WBAN), it ensures interactions between all kind of sensors (e.g. temperature, pressure, displacement...) and actuators (e.g. smart prosthetic, insulin pump...), as well as interfacing the human body (e.g. brain-machine interfaces) \cite{rabaey2015brain}.
	
	Positioned as a platform augmenting human capabilities, including life support applications, the Human Intranet must operate faultlessly.
%	as a standalone solution.
	It relies on a robust architecture, capable of mitigating different network topologies \cite{moin2017optimized} enabling high throughput and low latency \cite{englehart2003robust}, \cite{smith2010determining}, while being power efficient.

%	The Human Intranet is meant to act as a platform augmenting human capabilities.  and body dynamics [ref?]
	
%	Given the critical applications such as life support systems or smart prosthetics are falling under the human intranet umbrella, the network must also operate faultlessly as a standalone system.
%	The implementation of such a solution has to rely on a robust architecture, capable of mitigating the different network topology (objective-driven) \cite{moin2017optimized} and body dynamics, offering high throughput [X] and low latency [X] while being low power.
	The existing IEEE 802.15.6 standard \cite{ieee2008ieee} that cover such a network is good in terms communication reliability and security but has blocking limitations in its implementation: 1- or 2-hop network topology and heavy synchronization requirements limit its flexibility and significantly impact its efficiency.
%	Developed with that intent, the IEEE 802.15.6 standard \cite{ieee2008ieee} suffers severe weaknesses in its implementation: 1- or 2-hop network topology and heavy synchronization requirements limiting flexibility, both significantly impacting its efficiency.
%	 in terms of network topology (1- or 2-hop), heavy synchronization requirements limiting flexibility, significantly impacting its efficiency.
	
%	This is my introduction. The following topics must be addressed here: BAN, on-body and off-body communication. Specify the different network architecture possible for the application (star and mesh or both – Ali’s paper). Refer to IEEE standards such as 802.15.16 and other: not suitable, energy efficiency, collision...

	This paper introduces a new synchronization scheme based on the heartbeat to overcome those limitations. Highlighting promising results in \cite{li2009heartbeat}, a heartbeat-based synchronization scheme is proposed as a combination of puncture-based communication and dedicated hardware. Section~II details the synchronization principle. Section~III presents the models. The simulation results are provided in Section~IV along with a comparison with a duty-cycled architecture. Finally, Section~V concludes this paper.
	
%	Refer to X-MAC and H-MAC as initial idea for HB-based com. Regarding sync option, DuCy WRx/Rx is/are one but as exposed in (ref from Ove Edfors) may only be efficient with low traffic. In addition, how to deal with a non-deterministic latency. To overcome those limitations, a heartbeat based synchronization scheme is proposed.
	
%	The modeling and analysis of such a structure are covered in this 
%	The paper is organized as follow: 
%	regarding the efficiency of the heartbeat-based synchronization scheme in the Human Intranet context.

\section{Heartbeat-Based Synchronization Principle}\label{sec_HB_sync_princ}
	\subsection{Human Intranet architecture}
\begin{figure}[h]
			\centering
			\subfloat[]{\includegraphics[width=1.5in]{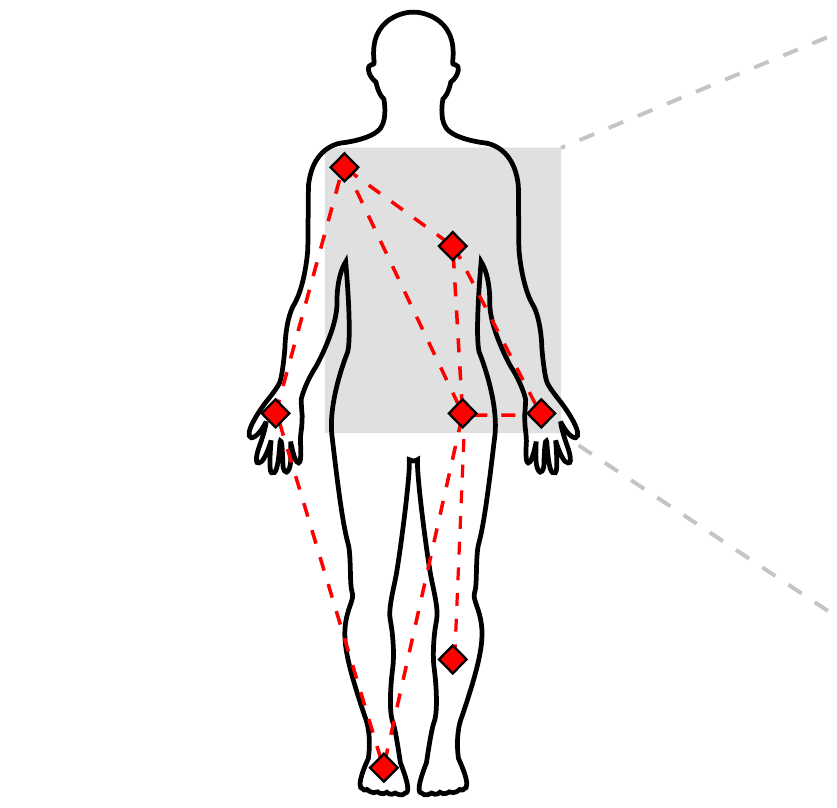}%
			\label{fig_HI_Network}}
			\hfil
			\subfloat[]{\includegraphics[width=1.5in]{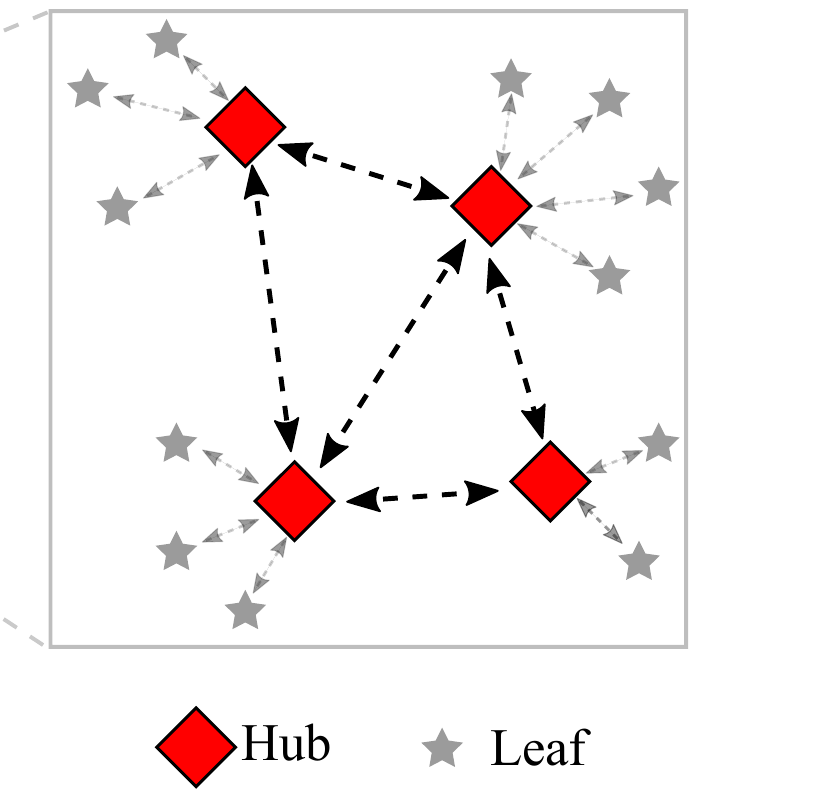}%
			\label{fig_Network_Archi}}
			\caption{Example of a Human Intranet Network.}
			\label{fig_HI_basics}
		\end{figure}

		The Human Intranet is meant to interconnect a wide range of wearable devices. They can be split into two categories: hubs and leaves (see Fig.~\ref{fig_HI_basics}\subref{fig_HI_Network}), based on their purpose (sensor or actuators), computing capabilities or access to energy \cite{movassaghi2014wireless}.
	
		The data traffic flows through the network as follows: The main traffic takes place between hubs in a bi-directional manner with the highest data rate achievable. It is the default situation. Each hub can exchange data with all the reachable hubs in its vicinity since no particular network topology is preferred (star, mesh or a combination of both). The leaves however, only communicate with one hub as shown in Fig.~\ref{fig_HI_basics}\subref{fig_Network_Archi}. The amount of data generated by a leaf is limited.
		
		Given the above architecture, a high level communication scheme is imagined where the leaves \enquote{puncture} the established communication between main nodes, to upload data to their respective hubs. This approach allows the system to take full advantage of the channel bandwidth available, enabling high data rate with simple communication scheme, offering a better energy efficiency \cite{abo2015modeling}.
		
	\subsection{Proposed synchronization scheme}
	
		The challenge of such a communication protocol lies in the nodes synchronization efficiency. The adequate coordination scheme optimizes three major parameters: the channel availability, the system power consumption and its latency. The channel availability, expressed as a percentage, is defined as the ratio between the hub-to-hub communication duration over the total considered communication period. Since the objective is to optimize the entire system, the power consumption calculation takes into account all nodes included in the analysis (i.e. at least a transmitter and a receiver). The latency represents the time elapsed from data availability to its transmission.
	%	(e.g. how long it takes to trigger an action when a given situation is identified).
		
\begin{figure}[h]
			\centering
			\includegraphics[width=3.2in]{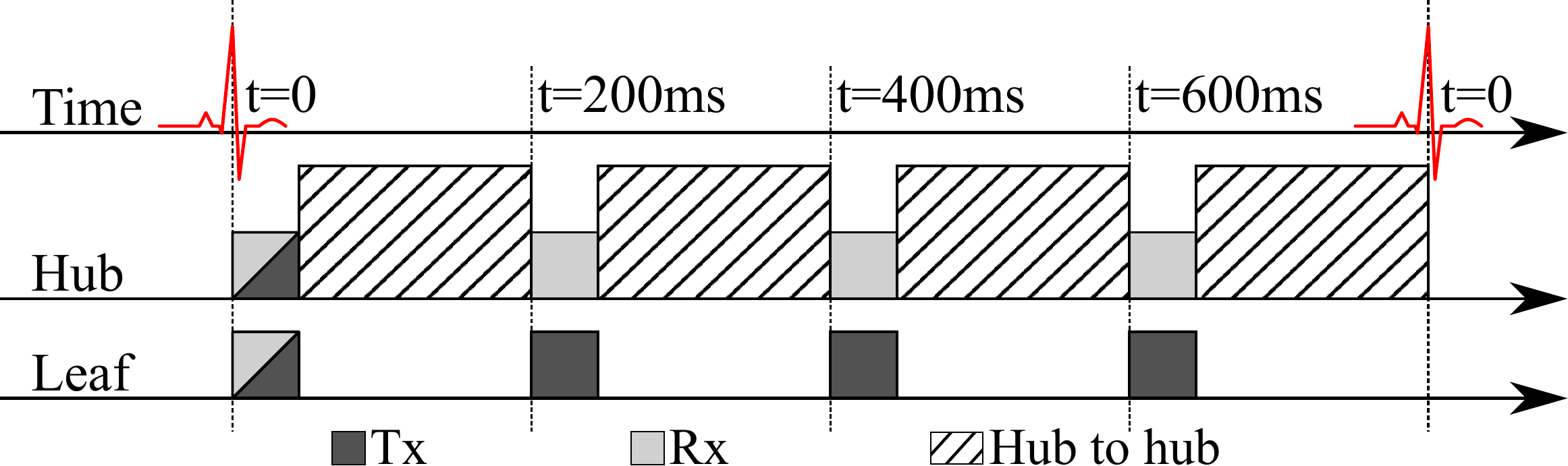}
			\caption{Superframe example with periodic puncturing and listening window.}
			\label{fig_Sufra_ex}
		\end{figure}

		The heartbeat-based synchronization requires all nodes to detect the heartbeat and use it as a time reference. A \enquote{superframe} is defined as the elapsed time between two heartbeats. Within this time frame, the communication between leaves and hubs is scheduled and periodically triggered. The synchronization scheme is depicted in Fig.~\ref{fig_Sufra_ex}.
		
		%All along the given time frame, called a \enquote{superframe} (i.e. between two heartbeats), the leaves communicate with their hub \enquote{on-appointment} only. An appointment is specified as an amount of time elapsed since the last heart beat and occurs recursively until the superframe ends. The synchronization scheme is depicted Fig.~\ref{fig_Sufra_ex}.
		
		A listening window is opened by the leaves once, at the beginning of each superframe. It provides a bi-directional communication capability allowing schedule update for instance.
		%For flexibility purposes, a listening window is open by the leaves at the beginning of each superframe, providing a bi-directional communication capability. This is the only moment where the hub can transmit data to a dedicated (or all) leaf until the following heart beat. The latter feature can be used for updating the appointment schedule for instance.
		
\begin{figure}[h]
			\centering
			\includegraphics[width=3.2in]{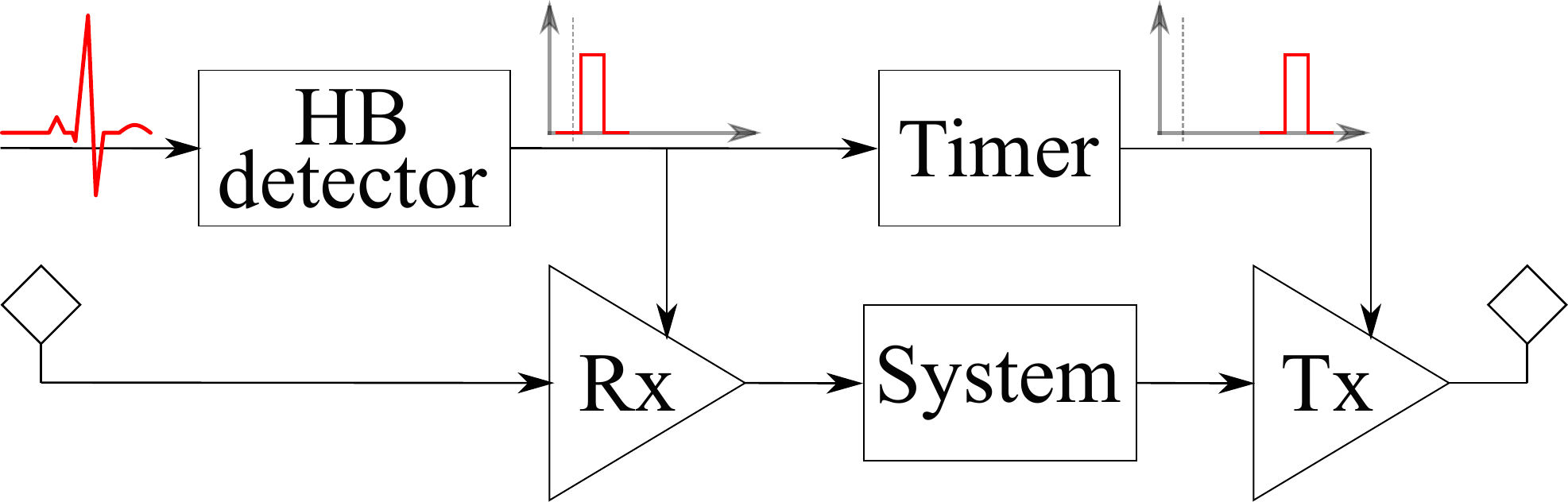}
			\caption{Synchronization functional block-diagram.}
			\label{fig_Sync_block_diag}
		\end{figure}

		Fig.~\ref{fig_Sync_block_diag} presents the heartbeat-based synchronization block-diagram. The  heartbeat detector (HB detector) translates the ECG signal into a digital output, resetting the timer. In parallel, the receiver is switched ON for a short period of time. The timer generates periodic ticks, triggering a transmission based on its counter auto-reload value. This scheme is repeated continuously until the next heartbeat. This study does not consider the set-up phase: the nodes schedule is already established.
		%Fig.~\ref{fig_Sync_block_diag} presents the heartbeat-based synchronization block diagram. The architecture embeds a heartbeat detector translating the ECG signal into a digital output. It resets the timer which generates periodic ticks based on its counter auto-reload setting. At the same time, the receiver is also switched on for a short period of time. Ultimately, when the timer goes off, a transmission is executed. The latter is repeated continuously until the next heartbeat.
		
		Many channel access methods \cite{doi:10.1002/0470861290.ch8} (e.g. TDMA) and synchronization implementations (e.g. RTC) already exist. However, exploiting the heartbeat as a synchronization signal has been rarely considered. To the author's knowledge the only previous related work is \cite{li2009heartbeat}. It offers unique advantages such as being available from the subject at no extra energy and transmission cost, and  adapting automatically to the wearer's physical activity. The heartbeat can be detected by various sensing modalities: electric, light, sound and pressure for instance.
	
\section{Hardware Architecture and Modeling}\label{sec_HW_modeling}
	This section details the functional blocks models ensuring the synchronization.
			
		\subsection{Heartbeat Detector}\label{subsec_HBd}
			The heartbeat detector identifies the peak (i.e. “R”) within the QRS complex \cite{zeng2015graphics}, not its shape. It translates the ECG signal into a digital output as implemented in \cite{da201458}. The heartbeat detector power consumption is noted as $P_{\mathit{hbd}}$.
			
			The main source of inaccuracy is the signal propagation delay between two nodes. From \cite{buchnerfast}, the signal propagation speed $v_{\mathit{HB}}$ is larger than $250$~m/s. Called \enquote{heartbeat skew}, the propagation delay $t_{\mathit{HB}}$ is only a function of the distance \textit{d} between the nodes: $t_{\mathit{HB}} \leq d/250$~s.
					
		\subsection{Timer}\label{subsec_Timer}
			The timer, embedding an oscillator and a counter, divides the superframe into sub-frames. It generates a signal when the elapsed time from the last heartbeat or timeout is equal to its internal setting. The timer is reset on each heartbeat, limiting inaccuracy accumulation over the running duration. Its power consumption is noted as $P_{\mathit{timer}}$.
			
			In terms of inaccuracy, three sources are identified: the oscillator frequency offset, the frequency drift and the accumulated random jitter.
			
			\subsubsection{Offset frequency}\label{subsubsec_Offset_freq}
				It specifies the deviation from the ideal frequency of oscillation. It is compensated by performing a one-time calibration, adjusting the counter auto-reload value. The offset frequency translates into a time uncertainty, $\Delta t_{counter}$, which eventually equals a single oscillator period.
				
%				only depends on the duration of one oscillation, is stated in (\ref{eq_offset_freq})
%				
%				\begin{equation}
%					\Delta t_{counter}=T_{osc}=\frac{1}{f_{osc}}
%					\label{eq_offset_freq}
%				\end{equation}
			
			\subsubsection{Frequency drift}\label{subsubsec_Freq_drift}
				It expresses the oscillation frequency variation due to close-in phase noise \cite{da2018understanding}. It is deterministic and bounded. It is calculated here as a ratio in ppm, $\mathit{Drift}_{osc}$, of the theoretical frequency of oscillation. The resulting inaccuracy, $\Delta t_{drift}$, is given in (\ref{eq_drift}).

				\begin{equation}
					\footnotesize{\Delta t_{drift}(t)=\mathit{Drift}_{osc}\cdot t}
					\label{eq_drift}
				\end{equation}
							
			\subsubsection{Accumulated random jitter}\label{subsubsec_Rand_jitter}
				It follows a Gaussian normal distribution. Its mean value is null since the drift is considered as a distinct parameter.
				
				Given the application (i.e. a timer), it is more relevant to analyze the accumulated random jitter over time. All cycles are  independent, and the accumulated random jitter also follows a Gaussian normal distribution \cite{zielinski2006estimation}, \cite{zielinski2009accumulated}, which variance $\sigma _N$ only depends on the number of oscillations $N$: $\sigma ^2_N=N \cdot \sigma^2$, where $\sigma$ is the jitter variance for one oscillation period.
				
				The inaccuracy due to the random jitter $\Delta t_{jitter}(t)$ in (\ref{eq_Rand_jit}) considers a window as large as $4 \cdot \sigma _{N}$, guaranteeing a covering probability higher than 99.993\%.
				
%				\begin{equation}
%					\begin{split}
%					\Delta t_{jitter}(t) & = 4 \cdot \sqrt{N} \cdot \sigma \\
%					& = 4 \cdot \sqrt{f_{osc} \cdot t} \cdot \sigma
%					\end{split}
%					\label{eq_Rand_jit}
%				\end{equation} 

				\begin{equation}
					\footnotesize{\Delta t_{jitter}(t) = 4 \cdot \sqrt{N} \cdot \sigma = 4 \cdot \sqrt{f_{osc} \cdot t} \cdot \sigma}
					\label{eq_Rand_jit}
				\end{equation}
		
		\subsection{Transmitter and Receiver}\label{subsec_Tx_Rx}
			In this study, the transmitter and receiver models are limited to their power consumption (no set-up time considered). The transmitter power consumption $P_\mathit{Tx}$ is a function of the link data rate $D_R$ times the Tx energy efficiency $\mathcal{E}_{\mathit{Tx}}$. The receiver energy consumption depends on the listening windows duration and the receiver power consumption $P_\mathit{Rx}$.
			
		\subsection{Metrics}\label{subsec_Metrics}
			The synchronization scheme efficiency is evaluated by the three metrics introduced in Section \ref{sec_HB_sync_princ}: channel availability, system power consumption and latency. Their analysis is conducted under two circumstances: the ideal and realistic case. The former does not take into account the loss while the latter includes the system nonidealities in terms of timing.
			
%			The former describes a situation with no loss while the latter takes into account the system nonidealities.
			
\begin{figure}[h]
				\centering
				\includegraphics[width=3.2in]{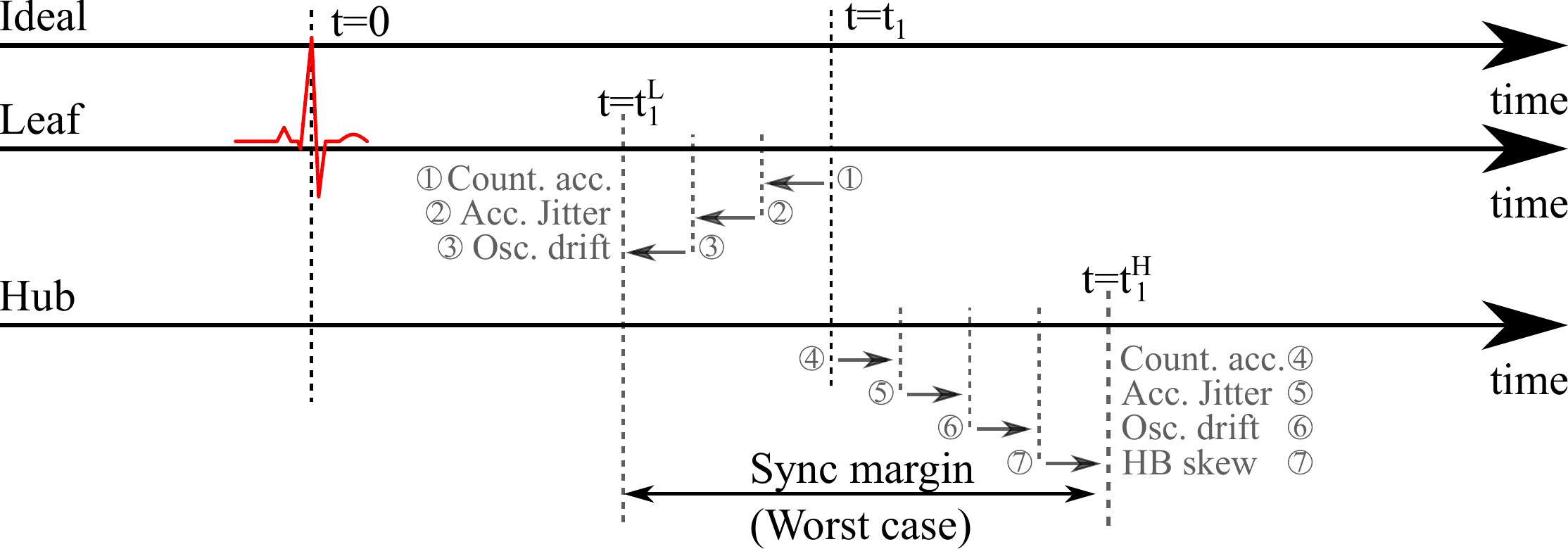}
				\caption{Synchronization margin example in realistic situation.}
				\label{fig_Sync_margin}
			\end{figure}

			In order to compensate for the timing inaccuracy within each node, the synchronization window must be properly chosen to ensure the communication. As depicted in Fig.~\ref{fig_Sync_margin}, a worst-case approach is followed. Maximum values of inaccuracy are considered with opposite consequences on Tx and Rx. The synchronization margin $M_S(t)$ taken per puncturing event is calculated in (\ref{eq_Sync_margin_per_punc}).
			
			\begin{equation}
				\footnotesize{M_S(t) = t_{\mathit{HB}} + 2 (\Delta t_{counter}+ \Delta t_{drift}(t) + \Delta t_{jitter}(t))}
				\label{eq_Sync_margin_per_punc}
			\end{equation}
			
			The synchronization margin increases over time. The total margin $M_{\mathit{tot}}$ taken for a node $i$ over the superframe duration is calculated in (\ref{eq_Tot_sync_marg}). It depends on the number of puncturing events $N_{p_i}$ for the given node.
			
			\begin{equation}
				\footnotesize{M_{\mathit{tot}}=\sum\limits_{j=1}^{\mathit{N_{p_i}}}\mathit{M_S(t_j)}}
				\label{eq_Tot_sync_marg}
			\end{equation}
			
			In case Rx is late compared to Tx (opposite of Fig.~\ref{fig_Sync_margin}), the total wait time, $t_{w_i}$, becomes twice the synchronization margin expressed in (\ref{eq_Sync_margin_per_punc}).
			
%			The realistic case necessary results in degraded performance, closer to a real implementation.
			
			\subsubsection{Channel availability}\label{subsubsec_CA}
				Under the ideal case, the total transmission time $T_{\mathit{Tx_i}}(t)$ per superframe period ($T$), equals the reception time. The ideal channel availability is calculated in (\ref{eq_CA_ideal}) for $n$ leaves connected to a hub.
				
				\begin{equation}
					\footnotesize{\mathit{CA_{ideal}}(t)=1-\sum\limits_{i=1}^{n}\frac{T_{\mathit{Tx_i}}(t)}{T}}
					\label{eq_CA_ideal}
				\end{equation}
				
				The realistic channel availability  $\mathit{CA_{real}}$ (\ref{eq_CA_real}), equals to $\mathit{CA_{ideal}}$ degraded by the total wait time per node.
				%, mandatory to ensure the synchronization between Tx and Rx (\ref{eq_CA_real}).
				
				\begin{equation}
					\footnotesize{\mathit{CA_{real}}(t)=\mathit{CA_{ideal}}(t)-\sum\limits_{i=1}^{n}\frac{t_{w_i}(t)}{T}}
					\label{eq_CA_real}
				\end{equation}
				
			\subsubsection{System power consumption}\label{subsubsec_Psys}
				Noted $P_{\mathit{ideal}}$ in the ideal case, it considers $n$ leaves and one hub (\ref{eq_Psys_ideal}).
				
%				\setlength{\arraycolsep}{0.0em}
%				\begin{eqnarray}
%				P_{\mathit{ideal}}&{}={}&(n+1)(P_{\mathit{hbd}}+P_{\mathit{timer}}) + (\mathcal{E}_{\mathit{Tx}}\nonumber\\
%				&&{+}\:\mathcal{E}_{\mathit{Rx}}) \cdot \frac{D_R}{T} \cdot \sum\limits_{i=1}^{n}T_{\mathit{Tx_i}}
%				\label{eq_Psys_ideal}
%				\end{eqnarray}
%				\setlength{\arraycolsep}{5pt}
				
				\begin{equation}
					\footnotesize{P_{\mathit{ideal}}(t) = (n+1)(P_{\mathit{hbd}}+P_{\mathit{timer}}) + (P_{\mathit{Tx}} + P_{\mathit{Rx}}) \sum\limits_{i=1}^{n}\frac{T_{\mathit{Tx_i}}(t)}{T}}
					\label{eq_Psys_ideal}
				\end{equation}
				
				The realistic system power consumption, $P_{\mathit{real}}$, increases as a function of the additional time spent waiting for data from the leaves. Its expression is given in (\ref{eq_Psys_real}).
				
				\begin{equation}
				\footnotesize{P_{\mathit{real}}(t) = P_{\mathit{ideal}}(t) + P_{\mathit{Rx}} \cdot \sum\limits_{i=1}^{n}\frac{t_{w_i}(t)}{T}}
				\label{eq_Psys_real}
				\end{equation}
			
			\subsubsection{Latency}\label{subsubsec_Lat}
				Given the synchronization mechanism and margin taken on each puncturing events, the latency does not suffer additional coordination time.
		
\section{Simulation Results}\label{sec_Sim_res}
	\subsection{Heartbeat-based synchronization scheme}
		In order to provide clear simulation results, hypotheses are made, or numerical values are chosen issued from existing implementations. This study also aims to define the system requirements and their impact on the synchronization efficiency. Conservative values are considered.
		
		In our Human Intranet scenario, the leaf-hub distance is not longer than 15~cm (about 50~cm between hubs) leading to $t_{\mathit{HB}} \leq 600~\mu$s.
		
		The system parameters introduced above and needed for the synchronization analysis are listed in Table \ref{tab_system_param}.
		
		\begin{table}[h]
			\renewcommand{\arraystretch}{1.3}
			\caption{System Parameters}
			\label{tab_system_param}
			\centering
			\begin{tabular}{|c|c|c|c|}
				\hline
				\bfseries Description & \bfseries Symbol & \bfseries Value & \bfseries Unit\\
				\hline 
				Oscillator jitter variance & $\sigma$ & 1 & $\mu$s\\
				\hline 
				Oscillator drift & $\mathit{Drift}_{osc}$ & 500 & ppm\\
				\hline
				Heart beat detector power consumption & $P_{hbd}$ & 100 & nW\\
				\hline
				Timer power consumption \cite{jeong20155}& $P_{timer}$ & 100 & nW\\
				\hline
				Tx energy efficiency & $\mathcal{E}_{\mathit{Tx}}$ & 100 & nJ/b\\
				\hline
				Rx energy efficiency & $\mathcal{E}_{\mathit{Rx}}$ & 100 & nJ/b\\
				\hline
				Communication data rate & $D_R$ & 100 & kb/s\\
				\hline
				Data generation rate (from leaf) & $D_{\mathit{gen}}$ & 1 & kb/s\\
				\hline
				Wake-up beacon length & $\mathit{WB}$ & 16 & b\\
				\hline
			\end{tabular}
		\end{table}
	
\begin{figure}[h]
			\centering
			\includegraphics[width=3.1in]{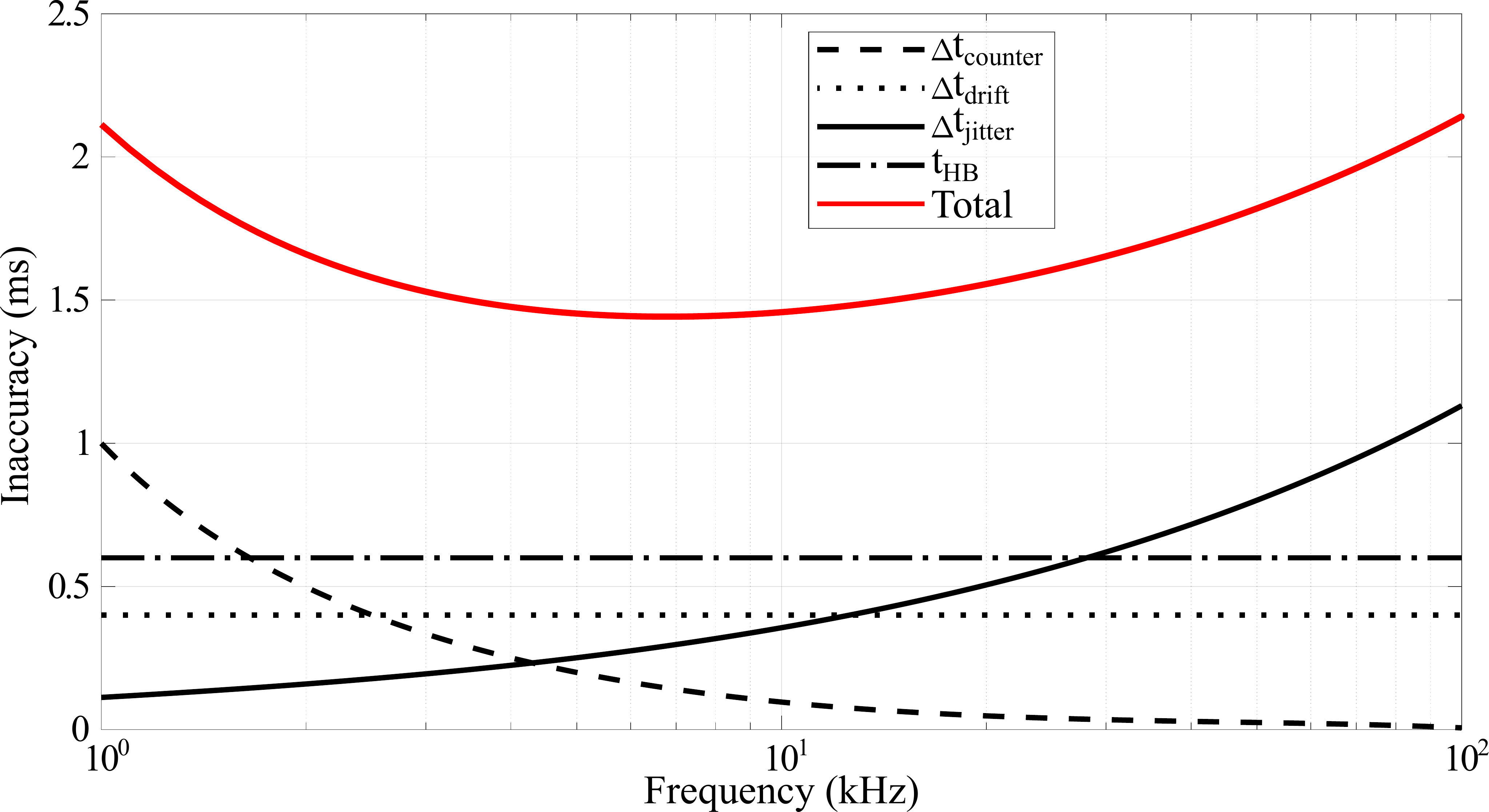}
			\caption{Total inaccuracy as a function of the oscillator frequency.}
			\label{fig_Osc_freq_simu}
		\end{figure}

		The oscillator frequency is selected $f_{osc}=10$~kHz. It is an optimal trade-off between timing granularity (i.e. $T_{osc}$) and drift over time, respectively dominating the inaccuracy at low and high frequencies. The overall timing inaccuracy and its components are plotted in Fig.~\ref{fig_Osc_freq_simu} for $T=800$~ms (inter-heartbeat duration at 75 beats per minute (bpm)).
		
\begin{figure}[h]
			\centering
			\subfloat[]{\includegraphics[width=1.7in]{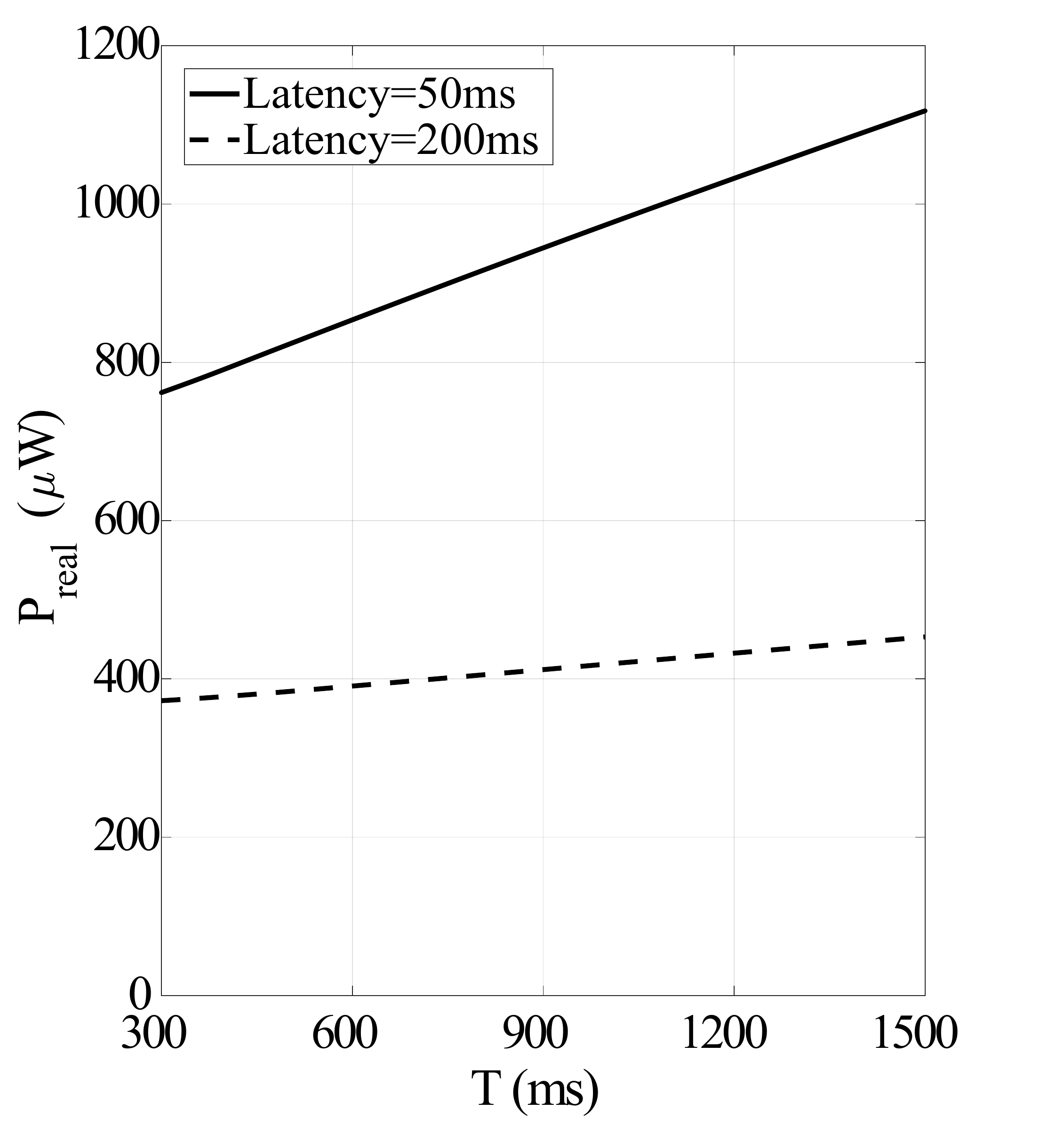}%
				\label{fig_HB_sync_Psys}}
			\hfil
			\subfloat[]{\includegraphics[width=1.7in]{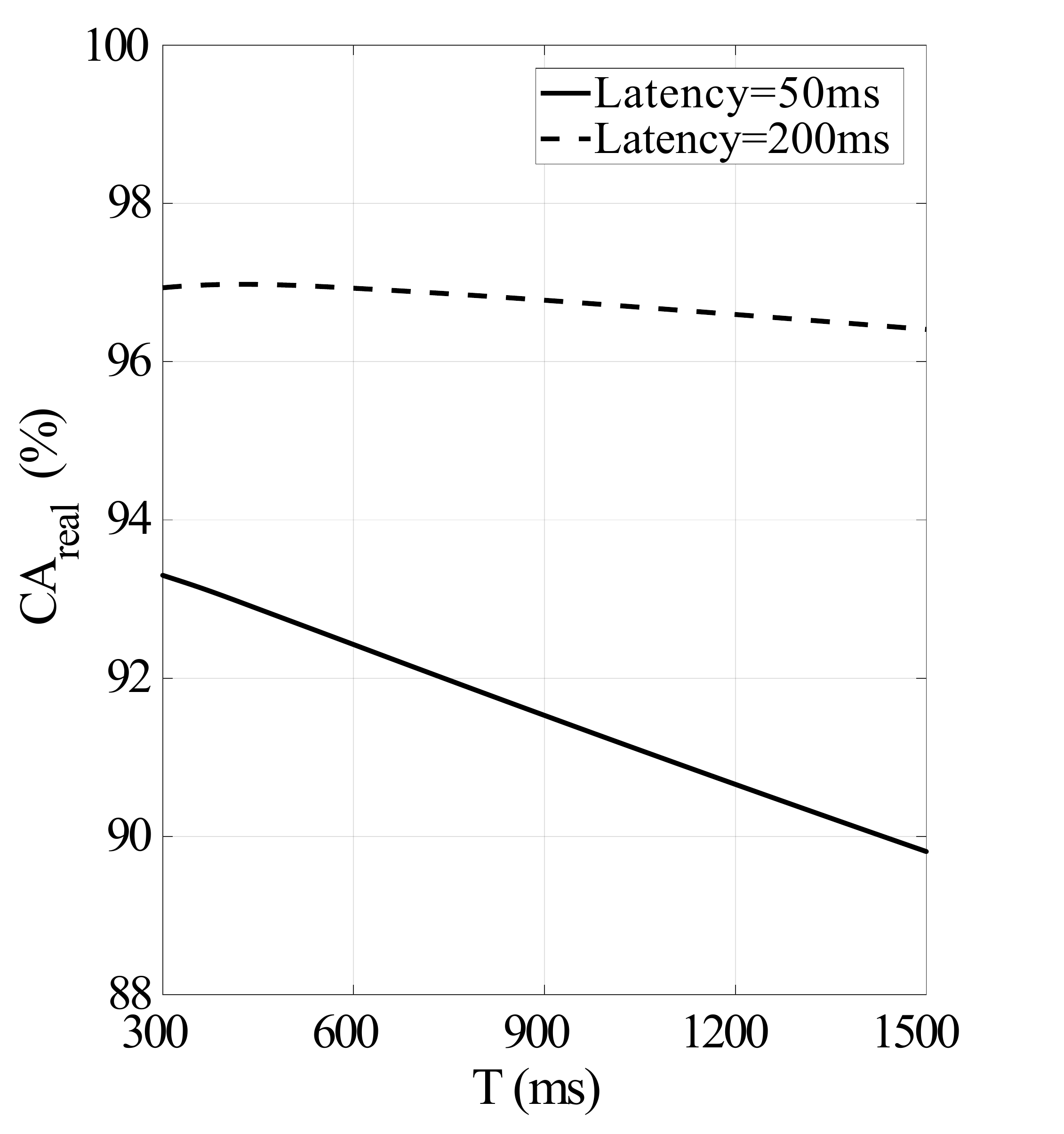}%
				\label{fig_HB_sync_CA}}
			\caption{System power consumption (a) and channel availability (b) in the realistic case for two latency settings.}
			\label{fig_HB_sync_CA_Psys}
		\end{figure}

		Following (\ref{eq_CA_real}) and (\ref{eq_Psys_real}), the channel availability and system power consumption are computed and plotted in Fig.~\ref{fig_HB_sync_CA_Psys}, in a 2-node configuration for two latency values: 50~ms and 200~ms.
		
		The impact of latency on channel availability illustrated in Fig.~\ref{fig_HB_sync_CA_Psys}\subref{fig_HB_sync_CA}, is inversely proportional to the duration between two consecutive uploads. At $T=800$~ms, $\mathit{CA_{real}}$ drops from 97\% to 92\% for latencies of 200~ms and 50~ms respectively. This observation is also applicable to $P_{\mathit{real}}$, since the additional term converting (\ref{eq_Psys_ideal}) into (\ref{eq_Psys_real}) is the same. For the given operating point, the consumption is more than doubled.
		
%		\begin{figure}[h]
%			\centering
%			\includegraphics[width=3in]{Figures/ISCAS_HBd_CA_P_2_Lat}
%			\caption{Channel availability (black) and system power consumption (grey) in the realistic case for two latency settings.}
%			\label{fig_HB_sync_CA_Psys}
%		\end{figure}
	
%		\begin{figure}[h]
%			\centering
%			\includegraphics[width=3.4in]{Figures/Total_wait_time}
%			\caption{Not sure if this figure is useful by any mean  but to show how to improve even more the system}
%			\label{fig_not_sure}
%		\end{figure}
	
	\subsection{Comparison with duty-cycled receivers}
		Receiver duty-cycling is one of the best recognized method for lowering a communication system power consumption [\cite{mazloum2017influence}].
		%As stated in the introduction, an efficient and well established synchronization scheme consists of duty-cycling a receiver.
		This method will be used as a comparison base. The duty-cycled system timing is described in \cite{mazloum2017influence}. Unlike the heartbeat-based solution, the nodes are not synchronized. The leaves randomly wake up, listening for a wake-up beacon from the hub. If a full wake-up message is detected, the transmission is initiated. They otherwise go back to sleep mode. The leaves listening windows last two wake-up beacons and one inter-wake-up beacon slots.
		
\begin{figure}[h]
			\centering
			\subfloat[]{\includegraphics[width=1.7in]{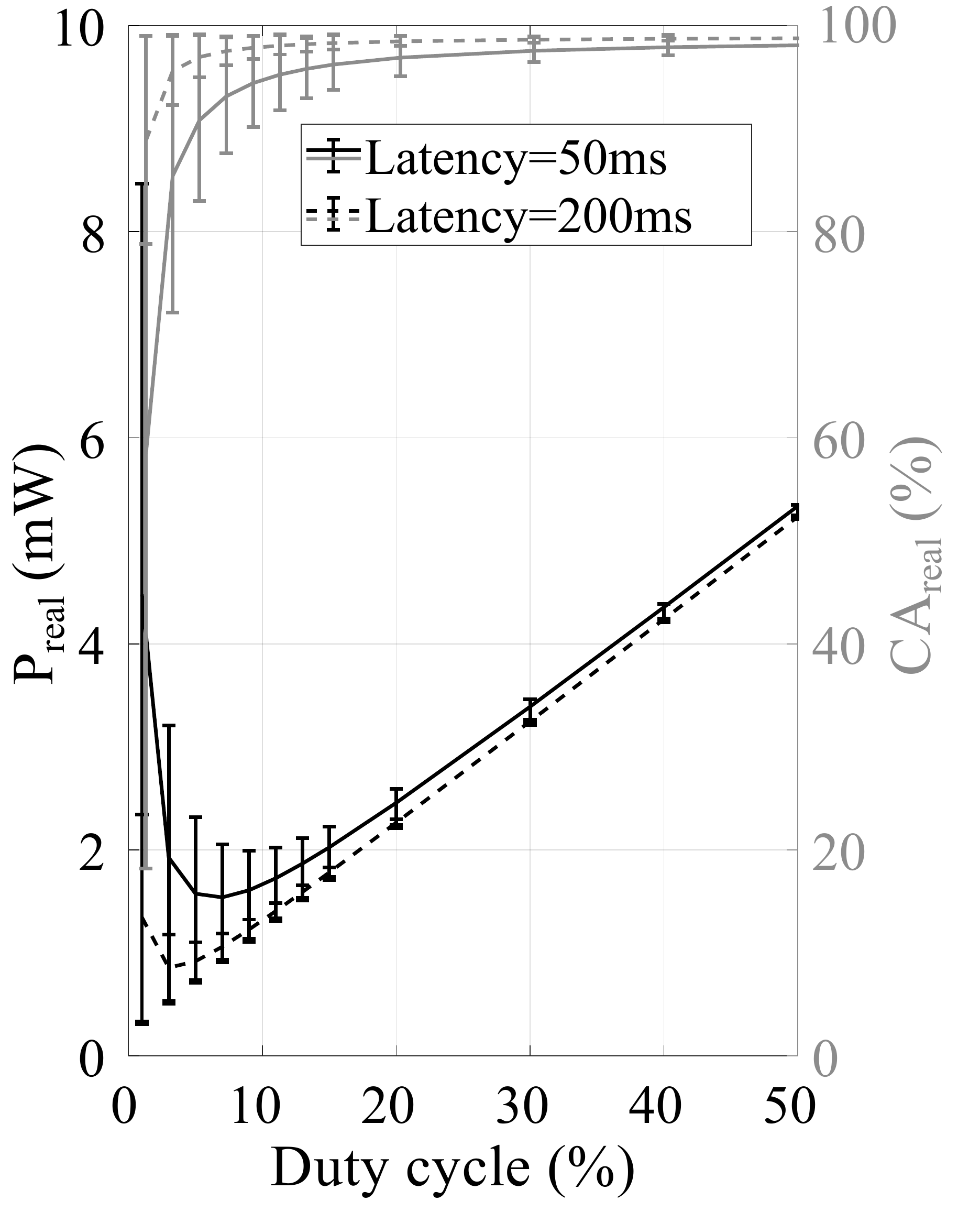}%
				\label{fig_DuCy_sync_CA_Psys}}
			\hfil
			\subfloat[]{\includegraphics[width=1.7in]{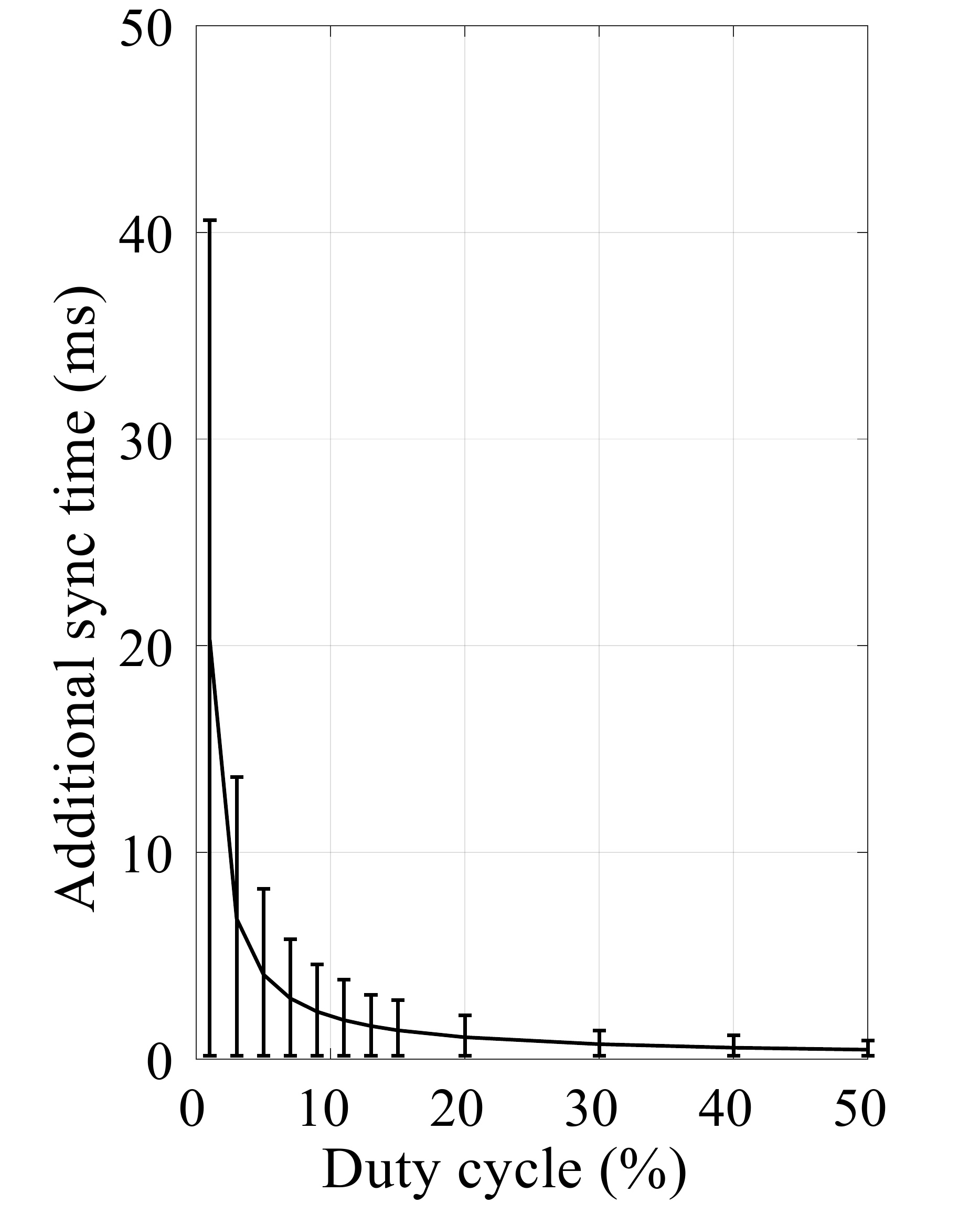}%
				\label{fig_Xtra_lat}}
			\caption{Channel availability, system power consumption and additional latency of a duty cycled architecture.}
			\label{fig_Ducy_sim_results}
		\end{figure}

		The performance of such a synchronization scheme is plotted in Fig.~\ref{fig_Ducy_sim_results} as a function of the duty-cycle ratio. The channel availability, system power consumption and latency are computed with the parameters from Table ~\ref{tab_system_param}. It is worth noting that the duty-cycle scheme relies on probabilities. To highlight this particularity, Fig.~\ref{fig_Ducy_sim_results} includes error bars, specifying the possible range of results around their mean.
		
		The duty-cycled architecture has an optimal power consumption for duty-cycles lower than 10\%, illustrated in Fig.~\ref{fig_Ducy_sim_results}\subref{fig_DuCy_sync_CA_Psys}. However, this minimum power consumption is higher than the heartbeat-based counterpart. Additionally, at this functioning point, the channel availability stays lower than the heartbeat-based equivalent.
		
%		The duty-cycled architecture has an optimal, higher than the heartbeat-based counterpart, power consumption illustrated in Fig.~\ref{fig_Ducy_sim_results}\subref{fig_DuCy_sync_CA_Psys}, for a duty cycled lower than 10\%. At this functioning point, the channel availability stays lower than the heartbeat-based equivalent.
		
%		\begin{figure}[h]
%			\centering
%			\includegraphics[width=3.4in]{Figures/ISCAS_DuCy_CA_P_2_Lat}
%			\caption{Channel availability and system power consumption of a duty cycled architecture case for two latency setting.}
%			\label{fig_DuCy_sync_CA_Psys}
%		\end{figure}
		
%		\begin{figure}[h]
%			\centering
%			\includegraphics[width=3.4in]{Figures/ISCAS_DuCy_Xtra_Lat}
%			\caption{Additional unpredictable latency after first wake-up beacon emission}
%			\label{fig_Xtra_lat}
%		\end{figure}
	
		The duty-cycle approach offers a better average channel availability for duty-cycle ratios higher than 10\%, for both latency requirements.
		
%		Summurized in Table \ref{tab_power_saving}, the heartbeat-based synchronization scheme improves the power consumption in regards to a duty cycled scheme.
		
%		\begin{table}[h]
%			\renewcommand{\arraystretch}{1.3}
%			\caption{System Parameters}
%			\label{tab_power_saving}
%			\centering
%			\begin{tabular}{|c|c|c|c|}
%				\hline
%				\bfseries Latency & \bfseries HBB scheme & \bfseries DuCy scheme & \bfseries Power saving\\
%				\hline 
%				50 (ms) & XX & XX & 26\%\\
%				\hline 
%				200 (ms) & YY & YY & 40\%\\
%				\hline
%			\end{tabular}
%		\end{table}
		
		The heartbeat-based scheme allows at least a 26\% and a 40\% power consumption saving while uploading data every 50~ms and 200~ms respectively, in a worst-to-best comparison (max of Fig.~\ref{fig_HB_sync_CA_Psys}\subref{fig_HB_sync_Psys}, min of Fig.~\ref{fig_Ducy_sim_results}\subref{fig_DuCy_sync_CA_Psys}).
		
		The last evaluation metric is the synchronization latency. Instantaneous in the heartbeat-based configuration due to a properly estimated listening window, the duty-cycled topology requires an additional synchronization time once the data is ready. Limited for duty-cycle ratios larger than 10\%, it is bounded but totally unpredictable otherwise (see Fig.~\ref{fig_Ducy_sim_results}\subref{fig_Xtra_lat}). This becomes problematic for demanding applications. It depends on the wake-up beacon length and increases exponentially for low duty-cycle ratios.
		
		The heartbeat-based synchronization scheme presents a more homogeneous (optimized combination of all three metrics) mean to initiate a communication between two nodes located on the body. It is a power efficient solution, optimizing the channel availability while offering tight control on the system latency, a crucial capability for critical applications.

\section{Conclusion}\label{sec_Conclusion}
A new synchronization scheme based on the heartbeat is introduced. It is composed of a timer, reset on each heartbeat, dividing the inter-heartbeat window (superframe) into sub-frames. This approach enables scheduled communication, optimizing the channel occupation and power consumption while offering a tight control on latency. A mathematical model, taking into account the hardware nonidealities is presented. It outlines the relationship between the hardware requirements and their impacts on the synchronization performance. Efficiency of the overall scheme is evaluated in a side-by-side comparison with a common duty-cycled receiver architecture. The heartbeat-based synchronization scheme provides more stable channel availability, a power consumption improved by at least 26\% without suffering latency uncertainty. In addition, all three parameters can be optimized together without suffering major trade-offs. The heartbeat-based synchronization scheme is a promising solution for an efficient Human Intranet implementation.

\newpage
\IEEEtriggeratref{10}
\bibliographystyle{IEEEtran}
\bibliography{IEEEabrv,ISCAS_References_Robin_Benarrouch}

\end{document}